\documentstyle[axodraw]{article}

\title{Non-universal gravitational couplings of neutrinos in matter
\footnote{This essay received an ``honorable mention'' in the 1999
Essay Competition of the Gravity Research Foundation}}
\author{Jos\'e F. Nieves\\
Laboratory of Theoretical Physics\\
Department of Physics, University of Puerto Rico \\
R\'{\i}o Piedras, Puerto Rico 00931-3343 \\[12pt]
Palash B. Pal\\
Saha Institute of Nuclear Physics\\
1/AF Bidhan-Nagar, Calcutta 700064, INDIA
}

\begin{document}
\maketitle 

\begin{abstract}
When neutrinos travel through a normal matter medium, the
electron neutrinos couple differently to gravity compared to the other
neutrinos, due to the presence of electrons in the medium and the absence
of the other charged leptons.  The matter-induced gravitational
couplings of the neutrinos under such conditions are calculated and 
their contribution to the neutrino index of refraction 
in the presence of a gravitational potential is determined. 
\end{abstract}

As is well known, the dispersion relation of a photon that propagates with
momentum $\vec K$ in a constant
gravitational potential $\phi^{ext}$ is
\begin{equation}
\label{photondisprel}
\omega_K = K(1 + 2\phi^{ext}) \,.
\end{equation}
This dispersion relation holds in fact not just for photons,
but for any massless particle.
In particular, it also holds for neutrinos, in the limit
in which their mass can be neglected. 
This fact is a consequence of the universality of
the gravitational interactions, which is embodied in
the Principle of Equivalence.  According to this principle, 
gravity couples to matter particles in a universal way; 
i.e., the graviton
field couples to the stress-energy tensor of the particles with
a coupling constant that is the same for all particle species.   

Since the gravitational contribution to the
neutrino dispersion relation is the same for all the
neutrino flavors, then it is not relevant
in the context of neutrino oscillation experiments.
That contribution yields a common factor in the evolution
equation and therefore it drops out of the formulas for the
flavor transition amplitudes.  
On the other hand, it has been
suggested by Gasperini\cite{gasperini},
and by Halprin and Leung\cite{hl}, that a small violation of
the equivalence principle, manifest as a difference in the couplings
of the various neutrinos to gravity, might explain the 
problems associated with the observations in neutrino experiments,
such as those involving solar neutrinos\cite{solarnus}.
These authors did not attempt to specify the source of this
violation of the equivalence principle, and instead assumed it 
as a working hypothesis.
Nevertheless, by the same token, precise measurements of various
observable quantities in these experiments can place
significant constraints on any deviations of the Equivalence
Principle in the neutrino sector.

This is an important point.
Due to its prominent role in the general theory of relativity, 
the Equivalence Principle has
been tested very accurately \cite{expts} in a variety of
experiments.  Those
tests ordinarily
involve particles that belong to the first generation of fermions,
of which normal matter is made.  It is much more difficult
to conceive analogous
tests involving the charged fermions in the other
generations, since they are unstable and therefore are not amenable for
performing precise tests of that kind with them. 
However, this may not be the case with neutrinos, 
which are believed to be stable particles and
therefore might provide the unique opportunity for testing the
equivalence principle across the generation gap,
as indicated above.

In this essay we point out that a breakdown
of the universality of the neutrino gravitational couplings
need not involve any fundamental violation
of the Equivalence Principle, but that it
indeed occurs when they propagate through a medium. This violation 
of the Equivalence Principle does not involve any
new physics beyond general relativity and the established 
properties of the standard particles and their
interactions.

The fact that the equivalence principle is violated in an
ambient thermal medium due to interactions with particles
in the heat bath is known\cite{dhr}. 
Using an illustrative model of thermal QED with only
photons in the background, it was demonstrated that the 
value of the inertial
mass of a charged fermion, which is obtained from pole of its
propagator, does not
coincide with the value obtained from its coupling to the
graviton in the weak-field limit.

Here we consider neutrinos
traveling through a background of normal matter, where electrons
and nucleons are present as well.
By calculating the weak interaction corrections to
the stress energy tensor of the 
neutrinos in the background, we show that the
lowest order gravitational couplings of the neutrinos are modified
in a non-universal way.
Besides explaining the origin of the breakdown of universality in this case,
the magnitude of the background-induced contribution
to the neutrino gravitational couplings is calculated,
the corrections to the neutrino dispersion relations 
in the presence of a gravitational potential are determined,
and some of their possible phenomenological consequences are mentioned.

In the linearized theory of gravity\cite{bd}, 
the graviton field is identified with the
tensor $h_{\lambda\rho}$, which is defined by writing
	\begin{eqnarray}
g_{\lambda\rho} = \eta_{\lambda\rho} + 2\kappa h_{\lambda\rho} \,,
\label{defkappa}
	\end{eqnarray}
where $g_{\lambda\rho}$ is the metric of the space-time and
$\eta_{\lambda\rho}$ is the flat metric. The dimensional coupling
$\kappa$ is related to the Newton constant by
	\begin{eqnarray}
\kappa = \sqrt{8\pi G} \,,
\label{kappa}
	\end{eqnarray}
so that the graviton field has the properly normalized kinetic term in
the linearized quantum theory of gravity.
In this theory the interaction Lagrangian of the graviton field
with other particles is
\begin{eqnarray}\label{Lffh}
{\cal L}_h = -\kappa h^{\lambda\rho} (x) \widehat T_{\lambda\rho} (x) \,, 
\end{eqnarray}
where $\widehat T_{\lambda\rho}(x)$ is the stress-energy-momentum
operator.  This implies that a fermion-fermion-graviton vertex
in a Feynman diagram is associated with the factor
$-i\kappa V_{\lambda\rho}(p,p')$, where $p$ and $p'$
denote the momenta of the incoming and the outgoing fermions, and
\begin{eqnarray}\label{Vmunu}
V_{\lambda\rho}^{(f)} (p,p') = \frac14 \Big[
\gamma_\lambda (p + p')_\rho + 
\gamma_\rho (p + p')_\lambda \Big]
- \frac12 \eta_{\lambda\rho}
\Big[ (\rlap/ p - m_f) + (\rlap/ p' - m_f) \Big] \,.
\end{eqnarray}
For the special case of chiral left-handed neutrinos,
\begin{eqnarray} \label{Vneutrino}
V_{\lambda\rho}^{(\nu)} (k,k') = \frac14 \Big[
\gamma_\lambda(k + k')_\rho + 
\gamma_\rho(k + k')_\lambda \Big] L
- \frac12 \eta_{\lambda\rho}
\Big[ \rlap/ k + \rlap/k \,' \Big] L \,,
\end{eqnarray}
where $L=\frac12 (1-\gamma_5)$.

\begin{figure}
\begin{center}
%
%
\begin{picture}(100,170)(-50,-30)
\Text(0,-30)[cb]{\large\bf (A)}
\ArrowLine(40,0)(0,0)
\Text(35,-10)[cr]{$\nu_i(k)$}
\ArrowLine(0,0)(-40,0)
\Text(-35,-10)[cl]{$\nu_i(k^\prime)$}
\Photon(0,0)(0,40){2}{6}
\Text(-4,20)[r]{$Z$}
\ArrowArc(0,60)(20,90,270)
\ArrowArc(0,60)(20,-90,90)
\Text(23,60)[l]{$f(p)$}
\Text(-23,60)[r]{$f(p - q)$}
\Photon(0,80)(0,120){2}{6}
\Photon(0,80)(0,120){-2}{6}
\end{picture}
%
%
\begin{picture}(100,170)(-50,-30)
\Text(0,-30)[cb]{\large\bf (B)}
\ArrowLine(40,0)(0,0)
\Text(35,-10)[cr]{$\nu_i(k)$}
\ArrowLine(0,0)(-40,0)
\Text(-35,-10)[cl]{$\nu_i(k^\prime)$}
\Photon(0,0)(0,40){2}{6}
\Text(-4,20)[r]{$Z$}
\ArrowArc(0,60)(20,-90,270)
\Text(0,85)[b]{$f(p)$}
\Photon(0,20)(40,20){2}{6}
\Photon(0,20)(40,20){-2}{6}
\end{picture}
%
%
\begin{picture}(100,170)(-50,-30)
\Text(0,-30)[cb]{\large\bf (C)}
\ArrowLine(40,0)(0,0)
\Text(35,-10)[cr]{$\nu_i(k)$}
\ArrowLine(0,0)(-40,0)
\Text(-35,-10)[cl]{$\nu_i(k^\prime)$}
\Photon(0,0)(0,40){2}{6}
\Text(-4,20)[r]{$Z$}
\ArrowArc(0,60)(20,-90,270)
\Text(0,85)[b]{$f(p)$}
\Photon(0,0)(40,20){2}{6}
\Photon(0,0)(40,20){-2}{6}
\end{picture}
%
%
\begin{picture}(100,170)(-50,-30)
\Text(0,-30)[cb]{\large\bf (D)}
\ArrowLine(40,0)(0,0)
\Text(35,-10)[cr]{$\nu_i(k)$}
\ArrowLine(0,0)(-40,0)
\Text(-35,-10)[cl]{$\nu_i(k^\prime)$}
\Photon(0,0)(0,40){2}{6}
\Text(-4,20)[r]{$Z$}
\ArrowArc(0,60)(20,-90,270)
\Text(0,85)[b]{$f(p)$}
\Photon(0,40)(40,30){2}{6}
\Photon(0,40)(40,30){-2}{6}
\end{picture}

\end{center}

\caption[]{\sf $Z$-exchange diagrams for the one-loop contribution
to the gravitational vertex of any neutrino flavor
($i = e,\mu,\tau$) in a background of electrons and
nucleons.  The
braided line represents the graviton.
\label{fig:zdiags}}

\end{figure}
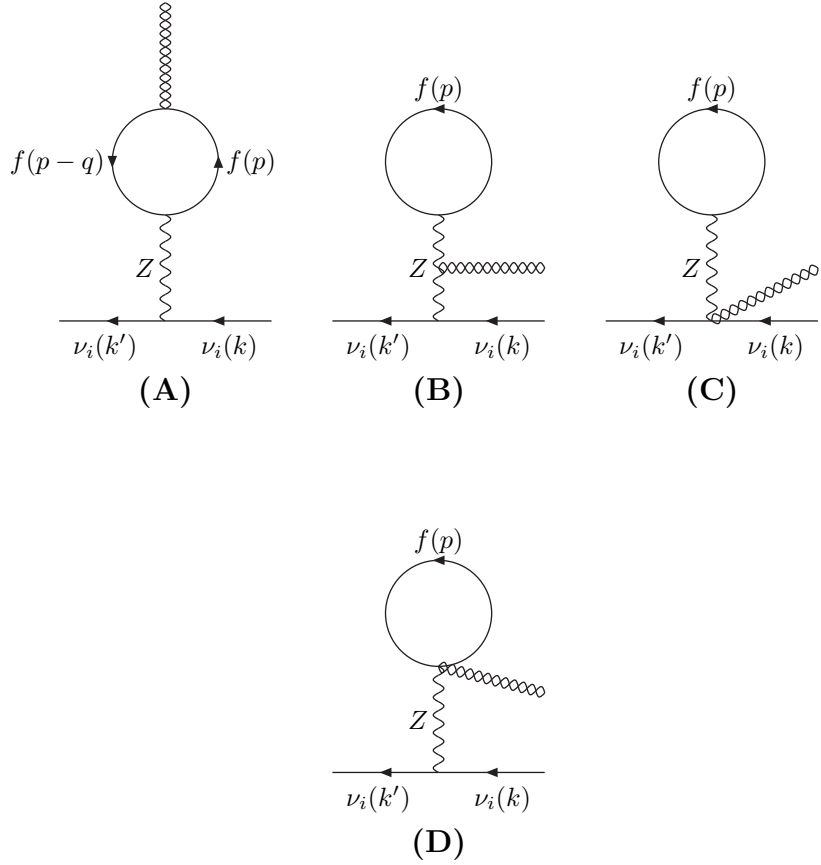
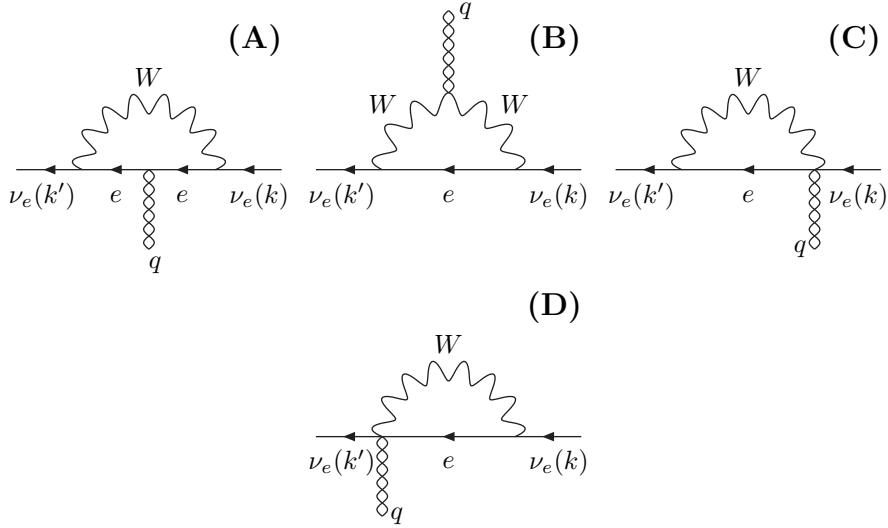
\begin{figure}
\begin{center}
%
%
\begin{picture}(110,100)(-55,-45)
\Text(40,55)[ct]{\large\bf (A)}
\ArrowLine(50,0)(25,0)
\Text(53,-10)[r]{$\nu_e(k)$}
\ArrowLine(25,0)(0,0)
\Text(12.5,-10)[c]{$e$}
\ArrowLine(0,0)(-25,0)
\Text(-12.5,-10)[c]{$e$}
\ArrowLine(-25,0)(-50,0)
\Text(-53,-10)[l]{$\nu_e(k')$}
\Photon(0,0)(0,-30){2}{3}
\Photon(0,0)(0,-30){-2}{3}
\Text(0,-33)[tl]{$q$}
\PhotonArc(0,0)(25,0,180){4}{7.5}
\Text(0,35)[c]{$W$}
\end{picture}
%
%
\begin{picture}(110,100)(-55,-45)
\Text(40,55)[ct]{\large\bf (B)}
\ArrowLine(50,0)(25,0)
\Text(53,-10)[r]{$\nu_e(k)$}
\ArrowLine(25,0)(-25,0)
\Text(0,-10)[c]{$e$}
\ArrowLine(-25,0)(-50,0)
\Text(-53,-10)[l]{$\nu_e(k')$}
\Photon(0,29)(0,60){2}{3}
\Photon(0,29)(0,60){-2}{3}
\Text(4,60)[l]{$q$}
\PhotonArc(0,0)(25,0,180){4}{6.5}
\Text(25,25)[c]{$W$}
\Text(-25,25)[c]{$W$}
\end{picture}
%
%
\begin{picture}(110,100)(-55,-45)
\Text(40,55)[ct]{\large\bf (C)}
\ArrowLine(50,0)(25,0)
\Text(53,-10)[r]{$\nu_e(k)$}
\ArrowLine(25,0)(-25,0)
\Text(0,-10)[c]{$e$}
\ArrowLine(-25,0)(-50,0)
\Text(-53,-10)[l]{$\nu_e(k')$}
\PhotonArc(0,0)(25,0,180){4}{7.5}
\Text(0,35)[c]{$W$}
\Photon(25,0)(25,-30){2}{3}
\Photon(25,0)(25,-30){-2}{3}
\Text(22,-30)[r]{$q$}
\end{picture}
%
%
%
\begin{picture}(110,100)(-55,-45)
\Text(40,55)[ct]{\large\bf (D)}
\ArrowLine(50,0)(25,0)
\Text(53,-10)[r]{$\nu_e(k)$}
\ArrowLine(25,0)(-25,0)
\Text(0,-10)[c]{$e$}
\ArrowLine(-25,0)(-50,0)
\Text(-53,-10)[l]{$\nu_e(k')$}
\PhotonArc(0,0)(25,0,180){4}{7.5}
\Text(0,35)[c]{$W$}
\Photon(-25,0)(-25,-30){2}{3}
\Photon(-25,0)(-25,-30){-2}{3}
\Text(-22,-30)[l]{$q$}
\end{picture}

\caption[]{\sf $W$ exchange diagrams for the one-loop contribution
to the $\nu_e$ gravitational vertex in a background of electrons.
\label{fig:wdiags}}

\end{center}
\end{figure}
When the particle propagates in a medium, 
this coupling is modified by interactions with
the background particles. In the language of Thermal Field
Theory, these modifications can be viewed as
corrections 
to $\widehat T_{\lambda\rho}$
arising from loop diagrams 
containing thermal particles\cite{npgravnu}. 
For neutrinos
traveling through a normal matter medium consisting of electrons and
nucleons, there are two kinds of diagrams, as shown in
Figs.~\ref{fig:zdiags} and \ref{fig:wdiags}. Among them,
Fig.~\ref{fig:zdiags} shows contributions arising from the neutral
current weak interactions, which are flavor independent. 
On the other hand, the diagrams shown in
Fig.~\ref{fig:wdiags} exist only for
$\nu_e$'s, and therefore contribute only to the
$\nu_e$-graviton vertex without affecting the graviton vertices of
$\nu_\mu$ or $\nu_\tau$. This is the source of the non-universal
coupling terms.

We now turn to the calculation of the diagrams.
The mass of the $W$ and the $Z$ bosons is taken to be much larger than any
other mass scale in the theory and we perform the
calculations to the leading order in the Fermi constant $G_F$. 
For the internal fermion lines
we use the propagators given by the
the real time formulation of Thermal Field Theory,
\begin{eqnarray}\label{SFf}
iS_F^{(f)}(p) = (\rlap/{p} + m_f)\left[
\frac{i}{p^2 - m_f^2 + i\epsilon} - 2\pi\delta(p^2 -
m_f^2)\eta_f(p)\right] \,, 
\end{eqnarray}
where
\begin{eqnarray}\label{etaf}
\eta_f(p) = \frac{\theta(p\cdot v)}{e^{\beta(p\cdot v - \mu_f)} + 1}
+ \frac{\theta(-p\cdot v)}{e^{-\beta(p\cdot v + \mu_f)} + 1}
\end{eqnarray}
with $\beta$ being the inverse temperature, $\mu_f$ the chemical
potential and $v^\mu$ the velocity four-vector of the medium. We will
work in the rest frame of the medium, where 
$v^\mu = (1,\vec 0)$.

The diagrams shown in
Fig.~\ref{fig:zdiags}A and
\ref{fig:wdiags}A have only one $W$ or $Z$ propagator
and hence give a correction of order $G_F$ to the
gravitational vertex.
Although the diagrams shown in Figs.~\ref{fig:zdiags}B and 
\ref{fig:wdiags}B have two
weak gauge boson propagators, the graviton coupling 
to the $W$ and the $Z$ also has
a term which is proportional to the squared mass of the gauge boson,
and so these diagrams also contribute to order $G_F$. These couplings
can be derived by starting from the general co-ordinate invariant form
of the pure $W$ and $Z$ Lagrangians and
keeping only the mass terms
\begin{eqnarray}
{\cal L}^{(W,Z)}_g = \sqrt{-\tt g} \; g^{\lambda\rho} \left[
M_W^2 W^\ast_\lambda W_\rho +
\frac12 M_Z^2 Z_\lambda Z_\rho \right] \,,
\end{eqnarray}
where ${\tt g}={\rm det}\, (g_{\lambda\rho})$. 
To first order in $\kappa$, this yields the graviton coupling term
\begin{eqnarray}
{\cal L}^{(W,Z)}_h = \kappa h^{\lambda\rho} a'_{\lambda\rho\alpha\beta}
\left[ M_W^2 W^\alpha W^{\ast\beta} + \frac12 M_Z^2 Z^\alpha Z^\beta
\right]  \,,
\end{eqnarray}
where
\begin{eqnarray}
a'_{\lambda\rho\alpha\beta} = \eta_{\lambda\rho} \eta_{\alpha\beta}-
(\eta_{\lambda\alpha}\eta_{\rho\beta} 
+ \eta_{\lambda\beta}\eta_{\rho\alpha}) \,.
\end{eqnarray}

In addition, there are contact couplings where the graviton couples to
a vertex of non-gravitational fields. To obtain them we start with the
general co-ordinate invariant form of the neutral current coupling,
\begin{eqnarray}
{\cal L}^{(nc)}_g = \sqrt{-\tt g} \Bigg[ -\frac{g}{2\cos\theta_W}
\Big( \sum_f \overline f
\gamma^a (X_f + Y_f \gamma_5) f \Big) \Bigg] v_a{}^\lambda
Z_\lambda \,,  
\end{eqnarray}
where $v_a{}^\lambda$ are the vierbeins, $X_f$ and $Y_f$ are the
vector and the axial coupling with the gauge boson of the fermion $f$,
and $g$ and $\theta_W$ are the usual parameters of the standard
electroweak model. Substituting
$v_a{}^\lambda = \eta_a{}^\lambda - \kappa h_a{}^\lambda\,,$
we get, to first order in $\kappa$, 
\begin{eqnarray}
{\cal L}^{(nc)}_h = -\kappa\frac{g}{2\cos\theta_W} h^{\lambda\rho}
Z^\beta a_{\lambda\rho\alpha\beta}  
\sum_f \overline f \gamma^\alpha (X_f + Y_f \gamma_5) f
\,,
\end{eqnarray}
where
\begin{eqnarray}
\label{symmform}
a_{\lambda\rho\alpha\beta} 
& = & \eta_{\lambda\rho}\eta_{\alpha\beta} - \frac{1}{2}(
\eta_{\lambda\alpha}\eta_{\rho\beta} + 
\eta_{\rho\alpha}\eta_{\lambda\beta}) \,.
\end{eqnarray}
The charged current couplings with the $W$-boson can be similarly derived.

We now calculate the amplitudes corresponding to the
various diagrams.  The
neutral current diagrams of Fig.~\ref{fig:zdiags} yield the
following background-dependent contributions to the vertex 
function\cite{footnote}
\begin{eqnarray}
\Gamma^{(\ref{fig:zdiags}A)}_{\lambda\rho} & = &
\Lambda^{(Z)}_{\lambda\rho} 
- b^{(Z)} \rlap/ v L\eta_{\lambda\rho} \,,\nonumber\\
\Gamma^{(\ref{fig:zdiags}B)}_{\lambda\rho} & = & -b^{(Z)}
a^\prime_{\lambda\rho\alpha\beta} \gamma^\alpha L v^\beta \,,\nonumber \\ 
\Gamma^{(\ref{fig:zdiags}C)}_{\lambda\rho} & = & b^{(Z)}
a_{\lambda\rho\alpha\beta} \gamma^\alpha L v^\beta \,,\nonumber \\
\Gamma^{(\ref{fig:zdiags}D)}_{\lambda\rho} & = & b^{(Z)}
a_{\rho\lambda\alpha\beta} \gamma^\alpha L v^\beta \,.
\end{eqnarray}
Here, $b^{(Z)}$ is a constant whose value is not important,
since the sum of the four diagrams
is simply $\Lambda^{(Z)}_{\lambda\rho}$, which is given by

	\begin{eqnarray}\label{LambdaZfinal}
\Lambda^{(Z)}_{\lambda\rho} = \surd 2 G_F
(\gamma^\alpha L)
\sum_{f}\int\frac{d^3p}{2E_f(2\pi)^3}
\Bigg\{X_f(f_f - f_{\overline f}) \left[
\frac{N^{(1)}_{\lambda\rho\alpha}(p_f,q)}{q^2 - 2p_f\cdot q} + 
(q\rightarrow -q)\right] \nonumber\\
- Y_f(f_f + f_{\overline f})\left[
\frac{N^{(2)}_{\lambda\rho\alpha}(p_f,q)}{q^2 - 2p_f\cdot q} -
(q\rightarrow -q) \right] \Bigg\} \,,
	\end{eqnarray}
where $f_f$ and $f_{\overline f}$ are the momentum
distribution functions for
particles and antiparticles, and
\begin{eqnarray}\label{N}
N^{(1)}_{\lambda\rho\alpha}(p,q) & \equiv &
(2p - q)_\lambda \big[2p_\rho p_\alpha
- (p_\alpha q_\rho + q_\alpha p_\rho) + (p\cdot q)\eta_{\alpha\rho}
\big] 
+ (\lambda\leftrightarrow\rho) \nonumber\\
N^{(2)}_{\lambda\rho\alpha}(p,q) & \equiv &
(2p - q)_\lambda i\epsilon_{\rho\alpha\beta\sigma}q^\beta p^\sigma +
(\lambda\leftrightarrow\rho) \,.
\end{eqnarray}
The same procedure for the $W$ diagrams yields
\begin{eqnarray}
\Gamma^{(\ref{fig:wdiags}A)}_{\lambda\rho} & = &
\Lambda^{(W)}_{\lambda\rho} -  
b_e\rlap/ v L\eta_{\lambda\rho} \nonumber\\
\Gamma^{(\ref{fig:wdiags}B)}_{\lambda\rho} & = &
b_e[-\eta_{\lambda\rho}\rlap/ v - \frac{1}{2} 
(\gamma_\lambda\rlap/ v\gamma_\rho + \gamma_\rho\rlap/
v\gamma_\lambda)]L\nonumber\\ 
\Gamma^{(\ref{fig:wdiags}C)}_{\lambda\rho} =
\Gamma^{(\ref{fig:wdiags}D)}_{\lambda\rho}  
& = & b_e[\eta_{\lambda\rho}\rlap/ v + \frac14
(\gamma_\lambda\rlap/ v\gamma_\rho + \gamma_\rho\rlap/
v\gamma_\lambda)]L \,. 
\end{eqnarray}
Their sum is just $\Lambda^{(W)}_{\lambda\rho}$,
which is obtained from Eq.\ (\ref{LambdaZfinal}) by making the
replacements 
\begin{eqnarray}\label{ZtoW}
X_e \to 1
\,, \quad Y_e \to -1 \,,
\end{eqnarray}
and setting all other $X_f,Y_f$ to zero.

Thus, the effective gravitational vertex function for the
neutrinos is
\begin{eqnarray}
\Gamma^{(\nu)}_{\lambda\rho} = V^{(\nu)}_{\lambda\rho} +
\Lambda_{\lambda\rho} \,,
\end{eqnarray}
where
\begin{eqnarray}\label{Lambda}
\Lambda_{\lambda\rho} = \left\{ \begin{array}{ll} 
\Lambda^{(W)}_{\lambda\rho} + \Lambda^{(Z)}_{\lambda\rho} & \mbox{for
$\nu_e$}\,, \\[12pt]
\Lambda^{(Z)}_{\lambda\rho} & \mbox{for $\nu_\mu,\nu_\tau$} \,.
\end{array} \right.
\end{eqnarray}

As these results clearly indicate,
the effective vertex $\Lambda_{\lambda\rho}$
is not the same for all the neutrino species.
As we have already argued, this is a manifestation
of the fact that the background is not flavor-symmetric.
One consequence of the non-universality
of the induced gravitational couplings is
that, in the presence of a static gravitational potential
$\phi^{\rm ext}$, the neutrino indices of refraction acquire
additional contributions proportional
to $\phi^{\rm ext}$ which are flavor-dependent.
To determine those we look at
the off-shell $\nu$-$\nu$ transition amplitude
in the presence of the external potential, which is given by
\begin{eqnarray}\label{Snunu}
S_{\nu\nu} = -2\pi i \kappa \,\delta(k^0 - k^{\prime 0})
\big(V^{(\nu)}_{\lambda\rho}(k,k') +
\Lambda_{\lambda\rho}(0,\vec{\cal Q}) \big)
h^{\lambda\rho}(\vec k' - \vec k) \,,
\end{eqnarray}
where
\begin{eqnarray}\label{hphirel}
h^{\lambda\rho}(\vec q \,') = \frac{1}{\kappa} \phi(\vec q \,') 
\big(2v^\lambda v^\rho - \eta^{\lambda\rho}\big) \,.
\end{eqnarray}
In Eq.\ (\ref{hphirel}) $\phi(\vec q \,')$ stands for 
the Fourier transform of $\phi^{\rm ext}$, which
for a homogeneous potential is given by
\begin{eqnarray}\label{phiconst}
\phi(\vec q\,') = (2\pi)^3\delta^3 (\vec q\,')
\phi^{\rm ext} \,.
\end{eqnarray}
Thus, substituting Eq.\ (\ref{phiconst}) into Eq.\ (\ref{Snunu}), we
obtain the following gravitational contribution to the 
neutrino self-energy induced by the background
\begin{eqnarray}
\label{SigmaG}
\Sigma_G & = & \phi^{\rm ext}
V^{(\nu)}_{\lambda\rho}(k,k)
\left(2v^\lambda v^\rho - \eta^{\lambda\rho}\right) + 
\phi^{\rm ext}
\Lambda_{\lambda\rho}(0,\vec{\cal Q}\rightarrow 0)
\left(2v^\lambda v^\rho - \eta^{\lambda\rho}\right) \nonumber\\
& = & \phi^{\rm ext} \left( \rlap/k + 2k\cdot v \rlap/v
\right) L + b_G\rlap{/}{v} \,,
\end{eqnarray}
where
\begin{eqnarray}\label{bGfinal}
b_G & = & \phi^{\rm ext} \sqrt{2}G_F \times
\left\{ \begin{array}{ll}
J_e +
\sum_f X_f J_f  & \mbox{for $\nu_e$}\,,\\
&\\
\sum_f X_f J_f   & \mbox{for
$\nu_\mu,\nu_\tau$\,,} 
\end{array}\right.  
\end{eqnarray}
and
\begin{eqnarray}\label{Jf}
J_f 
& = & -3(n_f - n_{\overline f}) + 4
\int\frac{d^3p}{(2\pi)^3 2E_f} \; \frac{d}{dE_f} \Big[ (2E_f^2 -
m_f^2)(f_f - f_{\bar f}) \Big] \,.
\end{eqnarray}
In Eq.\ (\ref{Jf}) $n_{f,\overline f}$
denote the number density of fermions and antifermions
in the background, respectively,
but the evaluation of the integral
can be made only if we make further assumptions about
the conditions of the background material. For example, 
\begin{eqnarray}
J_f = \left\{ \begin{array}{ll}
- \beta m_f n_f & \mbox{classical non-relativistic gas} \\
- 5n_f - m_f^2 \left( {3n_f\over \pi^4} \right)^{1/3} &
\mbox{degenerate non-relativistic gas} \\
- 5 (n_f-n_{\bar f}) & \mbox{ultra-relativistic gas} \,.
\end{array}
\right.
\end{eqnarray}

The presence of the $J_e$ term reflects the non-universality of neutrino
couplings with the gravitational field and has the consequence that
$\nu_e$ on one hand, and $\nu_{\mu,\tau}$ on the other,
have a different gravitational contribution to their
dispersion relation.  If we parametrize the dispersion
relation in terms of an index of refraction 
\begin{equation}
\label{refind}
{\cal N} \equiv \frac{K}{\omega_K} = 1 - 2\phi^{\rm ext} 
- \Delta{\cal N}_{\rm mat} - \Delta{\cal N}_G \,,
\end{equation}
where $\Delta{\cal N}_{\rm mat}$ stands for the usual Wolfenstein 
term\cite{msw,nr}, then Eq.\ (\ref{SigmaG}) implies that
\begin{equation}
\Delta{\cal N}_G = \frac{b_G}{K} \,.
\end{equation}

Therefore, in the presence of a static gravitational potential,
the matter-induced gravitational couplings 
lead to new contributions to the neutrino index
of refraction that could be relevant in the context
of matter-enhanced neutrino oscillations.
It is useful to notice that $\Delta{\cal N}_{\rm mat}$ and
$\Delta{\cal N}_G$ have a different dependence on the neutrino
coordinate as it propagates through the medium, a property
that could have distinctive implications.
Whether or not
these gravitational effects have observable consequences 
in specific contexts such as the supernova or the Solar
neutrino problem, remains an open question that deserves
further detailed study.  
However, we point out once more
that the calculations that we have presented, and the
results based on them, do not depend on any physical
assumption beyond those required by the standard model
of particle interactions and the linearized theory
gravity, including the question of whether or not the neutrinos
have a non-zero mass.
Hence, the effects that we have considered are
present at some level and it is
conceivable that they are detectable in some favorable
situations involving strong gravitational fields,
such as those that exist in the vicinity of active
galactic nuclei.  


\begin{thebibliography}{99}
\bibitem{footnote} The result of the evaluation of the individual
diagrams given in Ref.\ \cite{npgravnu} contained some errors that
originated in some incorrect signs that were adopted in the
gravitational vertices involved. The results quoted below are
the correct ones.  However, the formulas derived there for
the effective vertex functions $\Lambda^{(W,Z)}_{\rho\lambda}$,
as well as the final results for the gravitational contribution
to the neutrino self-energy and dispersion relations,
coincide with those obtained here.
\end{thebibliography}
\end{document}